\documentclass[preprint,amsmath,amssymb,aps,prb]{revtex4-1}
\usepackage{epsfig}
\usepackage{graphicx}
\usepackage{dcolumn}
\usepackage{bm}
\usepackage{indentfirst}
\usepackage{latexsym}
\begin{document}

\title{Defect-induced conductivity anisotropy in MoS$_2$ monolayers}

\author{Mahdi~Ghorbani-Asl,$^1$ Andrey~N.~Enyashin,$^{2, 3}$ Agnieszka~Kuc,$^1$ Gotthard~Seifert,$^2$ and~Thomas~Heine$^1$}
\email{t.heine@jacobs-university.de}
\affiliation{$^1$ School of Engineering and Science, Jacobs University Bremen, Campus Ring 1, 28759 Bremen, Germany\\ 
$^2$ Physical Chemistry, Technical University Dresden, Bergstr. 66b, 01062 Dresden, Germany\\
$^3$ Institute of Solid State Chemistry UB RAS, Pervomayskaya Str. 91, 620990 Ekaterinburg, Russia}
 
\date{}
\begin{abstract}
Various types of defects in MoS$_2$ monolayers and their influence on the electronic structure and transport properties have been studied using the Density-Functional based Tight-Binding method in conjunction with the Green's Function approach.
Intrinsic defects in MoS$_2$ monolayers significantly affect their electronic properties.
Even at low concentration they considerably alter the quantum conductance.
While the electron transport is practically isotropic in pristine MoS$_2$, strong anisotropy is observed in the presence of defects.
Localized mid-gap states are observed in semiconducting MoS$_2$ that do not contribute to the conductivity but direction-dependent scatter the current, and  that the conductivity is strongly reduced across line defects and selected grain boundary models.
\end{abstract}

\maketitle

\section{Introduction}
The rise of graphene\cite{Novoselov2005} launched the era of two-dimensional (2D) electronics, the manufacturing of electronic devices on substrates of one or few atomic layers thickness.
Graphene shows exceptional mechanical and electronic properties as well as spectacular physical phenomena, as for example massless Dirac fermions.\cite{Geim2007}
However, as in 3D electronics, the successful manufacturing of a variety of devices requires the combination of conducting, insulating and semi-conducting materials with tunable properties.
One class of 2D semiconductors and semimetals are transition-metal dichalcogenides (TMD).
Its most prominent representative, molybdenum disulphide (MoS$_2$), is a direct band gap semiconductor ($\Delta$ = 1.8 eV) in the monolayer (ML) form.\cite{Mak2010,Splendiani2010,Kuc2011}
Pioneering measurements of MoS$_2$-ML-based devices have shown that at room-temperature the mobility is about 200 cm$^2$~V~s$^{-1}$, when exfoliated onto the HfO$_2$ substrate, however, it decreases down to the 0.1--10 cm$^2$~V~s$^{-1}$ range if deposited on SiO$_2$.\cite{Radisavljevic2011}
Various electronic devices have been fabricated on MoS$_2$-ML, including thin film transistors,\cite{Radisavljevic2011, Pu2012, Kim2012} logical circuits,\cite{Wang2012} amplifiers\cite{Radisavljevic2012} and photodetectors.\cite{Lopez-Sanchez2013}
It has been shown that the electronic properties of MoS$_2$-ML can be easily tuned by doping,\cite{Ivanovskaya2006, Komsa2012, Dolui2013} bending\cite{Conley2013} or tube formation,\cite{Zibouche2012, Seifert2000} tensile strain\cite{Ghorbani2013} or intrinsic defects.\cite{Ataca2011, Zhou2013, Enyashin2013, VanderZande2013}

The chemical and structural integrity of MoS$_2$ depends on the manufacturing process.
Monolayers can be produced, following the top-down approach, from natural MoS$_2$ crystals by micromechanical exfoliation,\cite{Mak2010, Radisavljevic2011} intercalation based exfoliation,\cite{RamakrishnaMatte2010} or, on larger scale, by liquid-exfoliation techniques.\cite{Coleman2011}
On the other hand, chemical vapour deposition (CVD) is a bottom-up procedure and it provides a controllable growth of the material with the desired number of layers on the substrate of interest, e.g. on SiO$_2$\cite{Lee2012} or on graphene.\cite{Shi2012}

MoS$_2$-ML prepared in such different processes may contain numerous defects, including cationic or anionic vacancies, dislocations and grain boundaries.
Those defects significantly influence transport\cite{Lee2012} and optical properties\cite{Tongay2013} of these materials.
For example, it has been found that the maximum career mobility in CVD MoS$_2$ can be up to 0.02 cm$^2$~V$^{-1}$~s$^{-1}$,\cite{Lee2012}  while mechanically exfoliated ML showed a mobility of 0.1--10 cm$^2$~V$^{-1}$~s$^{-1}$.\cite{Novoselov2005, Radisavljevic2011}
Tongay et al.\cite{Tongay2013} showed that point defects lead to a new photoemission peak and enhancement in photoluminescence intensity of MoS$_2$-ML.
These effects were attributed to their trapping potential for free charge carriers and to localized excitons.

Defects may serve as means of engineering the MoS$_2$ properties, similarly as chemical impurities in semiconductor doping.
Zhou et al.\cite{Zhou2013} showed that S- and MoS$_3$-vacancies can be generated in CVD MoS$_2$-ML by extended electron irradiation.
This suggests that controlled defect-engineering allows tailoring -- even locally -- the electronic properties of MoS$_2$.

Structural defects in the TMD layers can appear in various types, such as point vacancies, grain boundaries, or topological defects.
The point vacancy is one of the native defects which have been investigated both in theory\cite{Komsa2012, Zhou2013a, Ma2011, Wei2012} and experiment.\cite{Zhou2013}
The recent experiment by Zhou et al.\cite{Zhou2013} showed that divacancies are only randomly observed, while monovacancies occur more frequently in MoS$_2$-ML.
Intrinsic defects can be created without elimination of atoms from the lattice, e.g.\ by performing Stone-Wales rotations and reconstructing intralayer bonds.\cite{Zhou2013}
First principles calculations by Zou et al.\cite{Zou2013} predicted that grain boundaries in MoS$_2$-ML can be formed as odd- or even-fold rings, depending on the rotational angle and stoichiometry, what has been been confirmed in experiment.\cite{VanderZande2013}
Line defects, suggested by Enyashin et al.,\cite{Enyashin2013} introduce a mirror plane into the MoS$_2$-ML, thus forming inversion domains.
Yong et al.\cite{Yong2008} showed that a finite atomic line of sulphur vacancies created on a MoS$_2$ surface can behave as a pseudo-ballistic wire for electron transport.

So far, direct measurement of the defect influence on the electronic structure and transport properties have been impossible because of substrate-induced local potential variations and contact resistances.
In order to fully understand and exploit defects in MoS$_2$-ML, we study here the electronic properties and the quantum transport of several structural defects using the density-functional based methods.
We will show that local defects introduce strongly localized mid-gap states in the electronic structure that act as scattering centers.
These scattering states do not open new transport channels, but they introduce high anisotropy in the quantum conductance.

\section{Methods}
\label{Sec:Methods}

All calculations have been carried out using the density-functional based tight-binding (DFTB) method\cite{Seifert1996, Oliveira2009} as implemented in the deMonNano code.\cite{deMonNano}
The structures of monolayers, that is, atomic positions and lattice vectors, have been fully optimized applying 3D periodic boundary conditions with a vacuum separation of 20~\AA\ perpendicular to the MLs.
The DFTB parameters for MoS$_2$-ML have been validated and reported earlier.\cite{Seifert2000, Kaplan-Ashiri2006}

\begin{figure}
\begin{center}
\includegraphics[scale=0.40,clip]{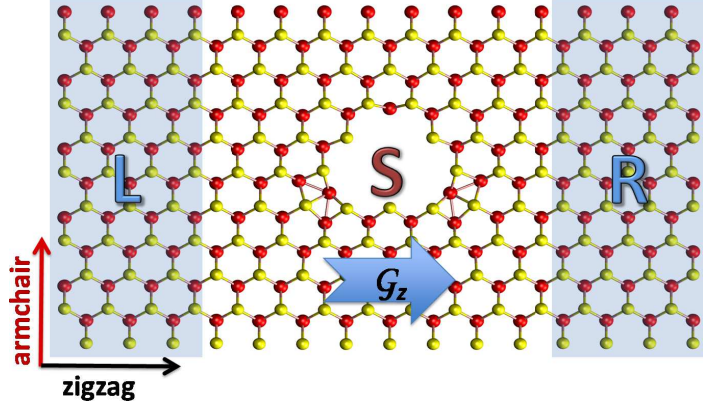}
\caption{\label{fig:1}(Color Online) Schematic representation of the electronic transport in the MoS$_2$-ML with a point defect. Left and right electrodes (L and R) that consist of semi-infinite ideal MoS$_2$-ML are highlighted. The scattering region (S) includes the defect. The transport direction is indicated by the arrow.}
\end{center}
\end{figure}

The coherent electronic transport calculations were carried out using the DFTB method in conjunction with the Green's function (GF) and the Landauer-B\" uttiker approach.\cite{Datta2005, DiCarlo2002}
Our in-house DFTB-GF software for quantum conductance has already been successfully applied to various nanostructures, including layered and tubular TMDs.\cite{Ghorbani2013, Miro2013, Ghorbani2013a}
The transport simulation setup consists of a finite defective MoS$_2$-ML as scattering region, which is connected to two semi-infinite ideal MoS$_2$-ML electrodes (Figure~\ref{fig:1}).
The selected scattering region is at least 28~\AA\ wide in order to prevent direct interaction between the electrodes.
The whole system is two-dimensional, and we apply in-plane periodic boundary conditions perpendicular to the transport direction. Thus, unphysical edge effects and out-of-plane periodicity are avoided.
Note that the electronic transport through the perfect monolayer represents the result for the bulk conductivity.
The quantum conductance ($\mathcal{G}$) was calculated at zero-bias following the Landauer-B\" uttiker formula,\cite{Landauer1970} where $\mathcal{G}$ is represented as:\cite{Fisher1981}
\begin{equation}
\mathcal{G}(E)=\frac{2e^{2}}{h}trace\left[\hat{G}^{\dagger}\hat{\,\Gamma}_{R}\hat{\, G}\hat{\,\Gamma}_{L}\right],
\end{equation}
where $\hat{G}$ denotes the total GF of the scattering region coupled to the electrodes and $\vphantom{} \hat{\Gamma}_{\alpha}=-2\mathrm{Im\mathit({\hat{\Sigma}_{alpha})}}$ is the broadening function, self-energies ($\hat{\Sigma}_{L, R}$) are calculated following the iterative self-consistent approach.\cite{Sancho1985}

\begin{figure}
\begin{center}
\includegraphics[scale=0.17,clip]{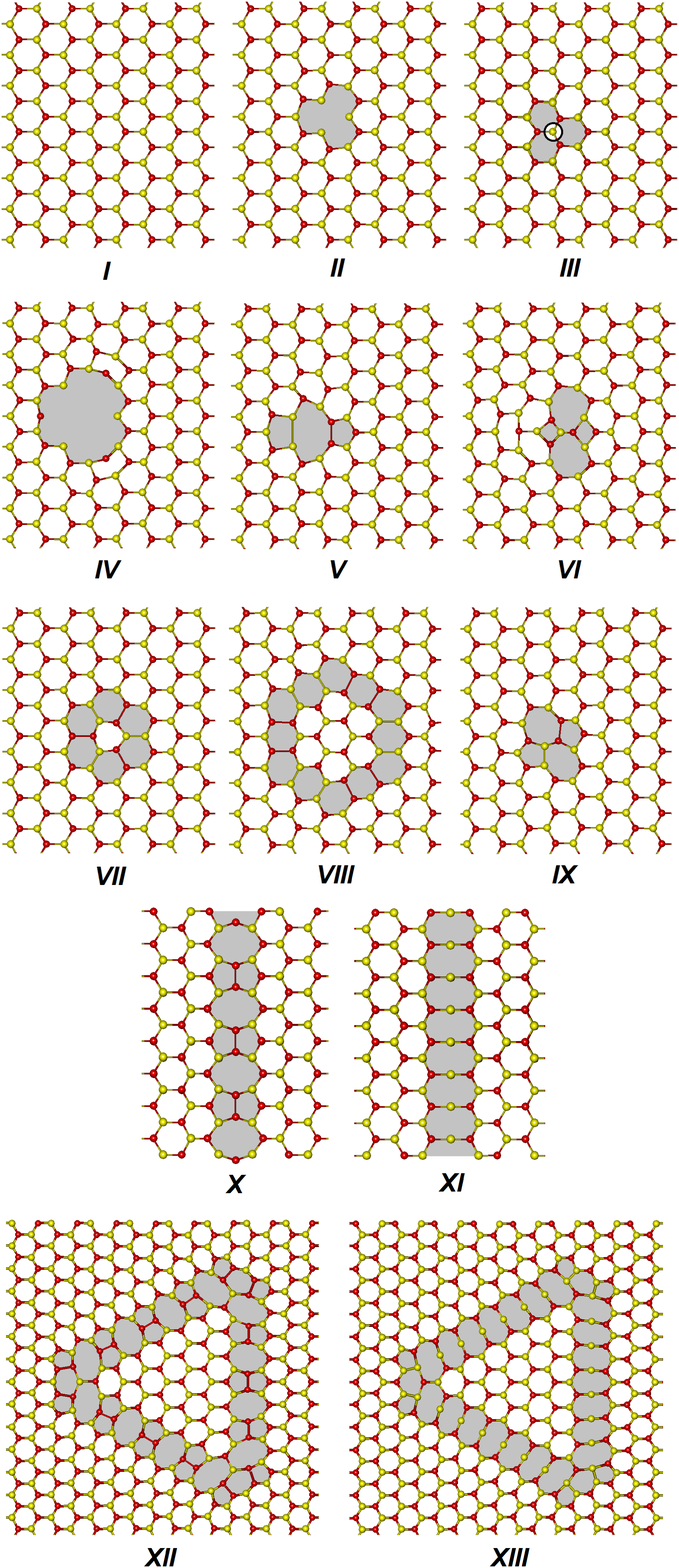}
\caption{\label{fig:2}(Color Online) Representation of local defects: Ideal (I) and defective MoS$_2$-ML containing point (II - IX), line (X - XI) defects and grain boundaries (XII - XIII). The defective areas are highlighted. Mo and S are shown in red and yellow, respectively. These figures show partial sections of the super cells used in the simulations.}
\end{center}
\end{figure}

The defective structures are shown in Figure~\ref{fig:2}.
Besides the pristine monolayer (I) we have studied three types of point defects, namely vacancies (II-V and XIII), add-atoms (VI), and Stone-Wales rearrangements (VII-IX).
Additionally, we have studied line defects formed from the vacancies (X, XI). 
In detail, we have considered non-stoichiometric single-atom vacancies of Mo and S atoms (II and III); multiple vacancies with dangling bonds (IV) or with reconstruction towards homonuclear bond formation (V); loops of line defects forming large triangular defects with homonuclear bond formation (XII and XIII).
The addition of one MoS$_2$ unit into the lattice (VI) causes rings of different oddity, such as "4-8" rings, preserving the alternation of chemical bonds.
A Stone-Wales rotation of a MoS$_2$ unit by 180$^{\circ}$ (VII and VIII) results in hexagonal rings with homonuclear bonds, while rotation by only 90$^{\circ}$ (IX) forms "5-7" rings, similar to the ones observed in graphene.\cite{Ma2009}
Line defects can be formed by vacancies of S or Mo atoms along zigzag direction (X and XI), resulting in the formation Mo-Mo and S-S dimer bonds, respectively.
Such systems have mirror symmetry along the defect lines, what imposes difficulties in the periodic model representation.

The point defects were simulated using the supercell approach, where the MoS$_2$ÐML was expanded to 90 Mo and 180 S atoms.
This supercell corresponds to the 5$\times$9 unit cells of the ideal lattice in rectangular representation.
The line defects were optimized using Mo$_{172}$S$_{344}$ supercells and, in order to maintain the in-plane 2D periodicity, both types of defects were present simultaneously in the optimization setup.
The transport calculations, however, were performed for each line defect separately, keeping the in-plane periodicity perpendicular to the transport direction.
Along the transport axis the scattering region was connected to semi-infinite electrodes and the whole system was treated using periodic boundary conditions.
Triangular island defects were represented using Mo$_{303}$S$_{586}$ and Mo$_{293}$S$_{606}$ supercells for S- and Mo-bridges, respectively.

As MoS$_2$-ML are produced under harsh conditions far from thermodynamic equilibrium, it can be assumed that a variety of defects are present in the samples.
Thus, we are not considering the thermodynamic stability of the defects, and refer the reader to recent studies on this subject.\cite{Enyashin2013, Enyashin2007}
All defects considered in this work are modeled by fully relaxed structures.
The geometry optimization did not reveal any considerable distortion of the layers and the defective structures preserved their integrity.

Note that the DFTB method, in the present implementations, does not account for the spin-orbit coupling (SOC) and therefore, this effect has not been considered in the present studies.
However, relativistic first-principles calculations including scalar relativistic effects and SO corrections showed that SOC in MoS$_2$-ML accounts for a large valence band splitting of about 145--148 meV, while the conduction band is affected in a much lower degree, by $\sim$3 meV band splitting.\cite{Zhu2011, Kormanyos2013}
At the same time, the effective masses in the valence change by about 5\%, while the effective masses in the conduction band are basically not affected by SOC.
The fundamental band gap changes by $\sim$ 50 meV.
Thus, the SOC effects will not significantly alter the results that are presented in the remainder of this work.

\section{Results and Discussion}

\begin{figure}
\begin{center}
\includegraphics[scale=0.60,clip]{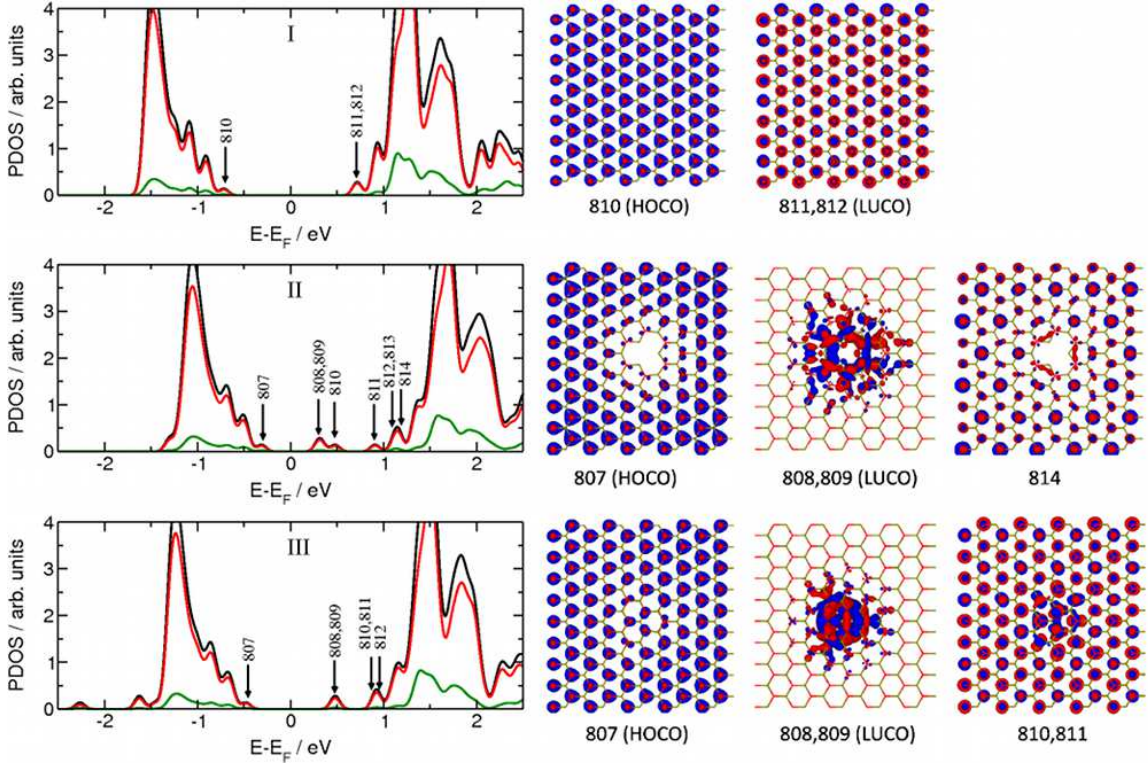}
\caption{\label{fig:3}(Color Online) (Left panel) Total (black), Mo-4$d$ (red) and S-3$p$ projected (green) densities of states (DOS) of selected defective MoS$_2$-ML (labels as in Figure~\ref{fig:2}).  (Right panel) Highest-occupied and lowest-unoccupied crystal orbitals (HOCO and LUCO), and delocalized conduction orbitals.}
\end{center}
\end{figure}

We have investigated the electronic structure of defective MoS$_2$-ML by calculating orbital-projected densities of state (PDOS).
The results were compared with the perfect MoS$_2$ system.
Crystal orbitals are visualized corresponding to the states close to the Fermi level (E$_F$) (Figure~\ref{fig:3}).
The electronic structure of MoS$_2$-ML suggests that the bottom of the conduction band is formed from empty Mo-4$d_{z^2}$ orbitals,\cite{Enyashin2013, Kuc2011} while the top of the valence band is composed of fully occupied $d_{xy}$ and $d_{x^2-y^2}$ orbitals, in agreement with the crystal-field splitting of trigonal prismatic systems.
In pristine MoS$_2$, the highest-occupied and lowest-unoccupied crystal orbitals (HOCO and LUCO) are delocalized and spread homogeneously throughout the system (Figure~\ref{fig:2} I).
The electronic band gap of MoS$_2$-ML obtained at the DFTB level of about 1.5 eV is smaller than that obtained from DFT\cite{Mak2010, Kuc2011, Splendiani2010} and experiment\cite{Mak2010} due to the deviations in geometry.
DFTB-estimated lattice vectors ($a$ = 3.32~\AA) are by 5\% larger than the experimental values ($a$ = 3.16~\AA),\cite{Wilson1969} and, as it has been discussed earlier, such a distortion in geometry causes the decrease in the band gap.\cite{Ghorbani2013}
However, these discrepancies should not alter general trends and conclusions drawn here, as the relative change in the electronic structure is not influenced.

As known from the literature,\cite{Komsa2012, Zhou2013a} defects result in significant changes of the electronic structure close to E$_F$ and introduce mid-gap states.
The mid-gap states are strongly localized in the vicinity of the defects and are mostly of 4$d$-Mo type, thus they act as scattering centers.
Although defect states reduce the band gap significantly, the scattering character will prevent opening any new conduction channels close to E$_F$.

In case of a single Mo vacancy (Figure~\ref{fig:3} II), the valence band maximum (HOCO) resembles the characteristics of perfect MoS$_2$-ML, while the edge of conduction band is formed by two individual mid-gap states (states 808/809 and 810 in Figure~\ref{fig:3} II), the first being the LUCO.
The next delocalized states are located at around 1.2 eV above E$_F$ (state 814).
A similar situation is observed for the single S vacancy (Figure~\ref{fig:3} III), with the LUCO (degenerated states 808/809) composed of single strongly localized states.
In this case, the next delocalized states are present only at about 1 eV above E$_F$ (810/811).

For larger point defects, where both types of elements are removed from the lattice, the electronic structure changes even stronger, with HOCO states being no longer delocalized.
In the case of Stone-Wales defects, the band gap reduces with the number of rotated bonds and introduces a larger number of mid-gap states.
At the same time, the HOCO becomes more localized (Figure S1 and S2 in Supporting Information).\cite{SI}

Considering the triangular domain structures (XII and XIII), which contain Mo-Mo and S-S line defects, the PDOS shows interesting characteristics.\cite{Enyashin2013}
The Mo-Mo bridges contribute primary with the 4$d$ to the HOCO and the LUCO.
These are localized states at about 2.5 eV below E$_F$, indicating the formation of 
strong Mo$-$Mo bonds.
In contrast, the S-S line defects form S-3$p$ states, which do not contribute to the PDOS close to E$_F$. These states can be found deep in the valence band region at about 3~eV below E$_F$.
The states in the vicinity of E$_F$ are, therefore, composed exclusively from the edge states of the MoS$_2$ domains.

The defect-induced variations in the electronic structure affect the electronic transport in the MoS$_2$-ML.
The transport through MoS$_2$ should be direction dependent due to the structural anisotropy of the system.
The pristine layer shows, however, very little anisotropy in the electron conductivity as reported earlier.\cite{Ghorbani2013}

The two extremes are transport along armchair ($\mathcal{G}_a$) or zigzag ($\mathcal{G}_z$) directions.
In order to ensure transferability of the results, we used a supercell with almost equal length and width along both transport directions (L$_a$ = 28.75~\AA\ and L$_z$ = 29.88~\AA). 

\begin{figure}
\begin{center}
\includegraphics[scale=0.40,clip]{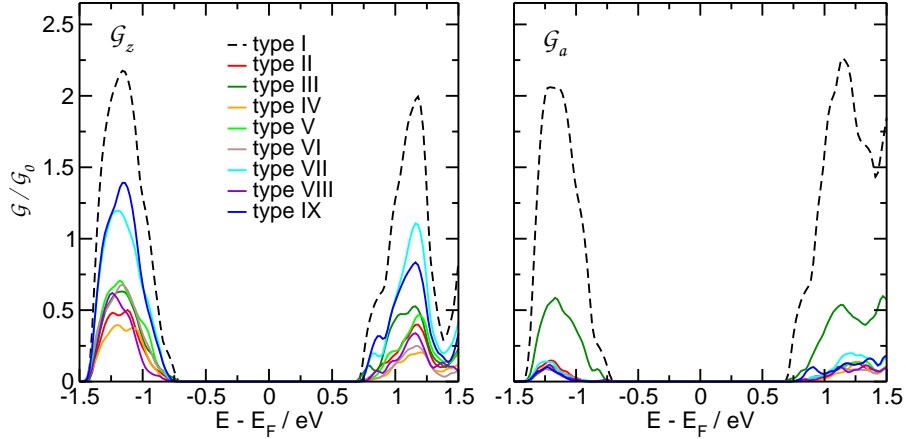}
\caption{\label{fig:4}(Color Online) Electron conductivity of MoS$_2$-ML with point defects. $\mathcal{G}_a$ (a) and $\mathcal{G}_z$ (b) denote electron conductivity along the armchair and zigzag direction, respectively. Labels are as in Figure~\ref{fig:2}.}
\end{center}
\end{figure}

Figure~\ref{fig:4} shows the electron conductivity of the MoS$_2$-ML in the pristine form and in the presence of various point defects along the armchair and zigzag directions.
The occurrence of defects reduces the conductivity (transmittance) in comparison with the pristine layer at 1.2 eV below and above E$_F$.
This is expected as the vacancy causes backscattering effects,\cite{Rutter2007, Deretzis2010} and, not surprisingly, the conductivity depends strongly on the type and concentration of the point defects.
Noteworthy, in contrast to the pristine ML, the electron conductivity of the defective systems becomes strongly direction dependent and the conductivity is suppressed much stronger along the armchair direction.
The only exception is the single S-vacancy, where the transport is rather direction independent.

The directional dependence of the conductance might arise from different transmission pathways\cite{Wang2009, Liu2013}  and electron hopping within defective parts of MoS$_2$-ML.\cite{Remskar2011}
To date, only grain boundaries have been studied in experiment, and our results are consistent with the results reported by the Heinz group.\cite{VanderZande2013}

In case of transport in armchair direction, the electron conductance of MoS$_2$ with one Mo-vacancy, corresponding to 1.11\% structural defects, is suppressed by 75\% compared with the pristine layer.
The single S vacancy (0.55\% structural defects) shows higher electron conductivity due to the electron injection directly to the conduction band.
For the Stone-Wales defects (VII and IX) the conductance is reduced by less than 50\% with respect to the pristine structure.

\begin{figure}
\begin{center}
\includegraphics[scale=0.40,clip]{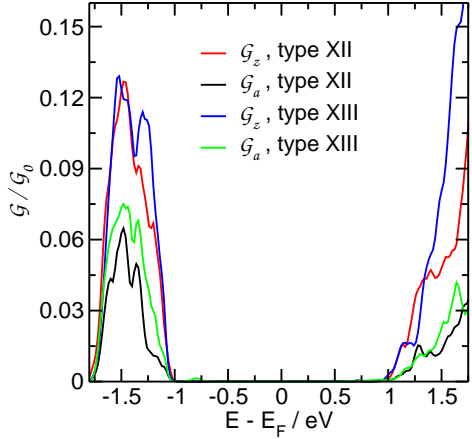}
\caption{\label{fig:5}(Color Online) Electron conductivity of MoS$_2$-ML with grain boundaries formed by inversion domains. $\mathcal{G}_a$ (a) and $\mathcal{G}_z$ (b) denote electron conductivity along the armchair and zigzag direction, respectively. Labels as in Figure~\ref{fig:2}.}
\end{center}
\end{figure}

Figure~\ref{fig:5} shows the transport properties of the MoS$_2$-ML with triangular grain boundaries along the armchair and zigzag directions.
It is very interesting to notice that $\mathcal{G}$ in this case does not depend on the type of the defect and similar values are obtained for Mo--Mo and S--S bridges.
The conductance is, however, strongly direction-dependent and again we observe that along the armchair lines it is more suppressed than along the zigzag ones.

Our results indicate that local defects introduce spurious minor conductance peaks close to E$_F$ (see Figure S3 in Supporting Information\cite{SI}).
Because these electronic states are strongly localized, they do not contribute to the overall quantum transport, as they cannot generate additional conducting channels for a specific energy window within the perfect semi-infinite electrodes.

\begin{figure}
\begin{center}
\includegraphics[scale=0.30,clip]{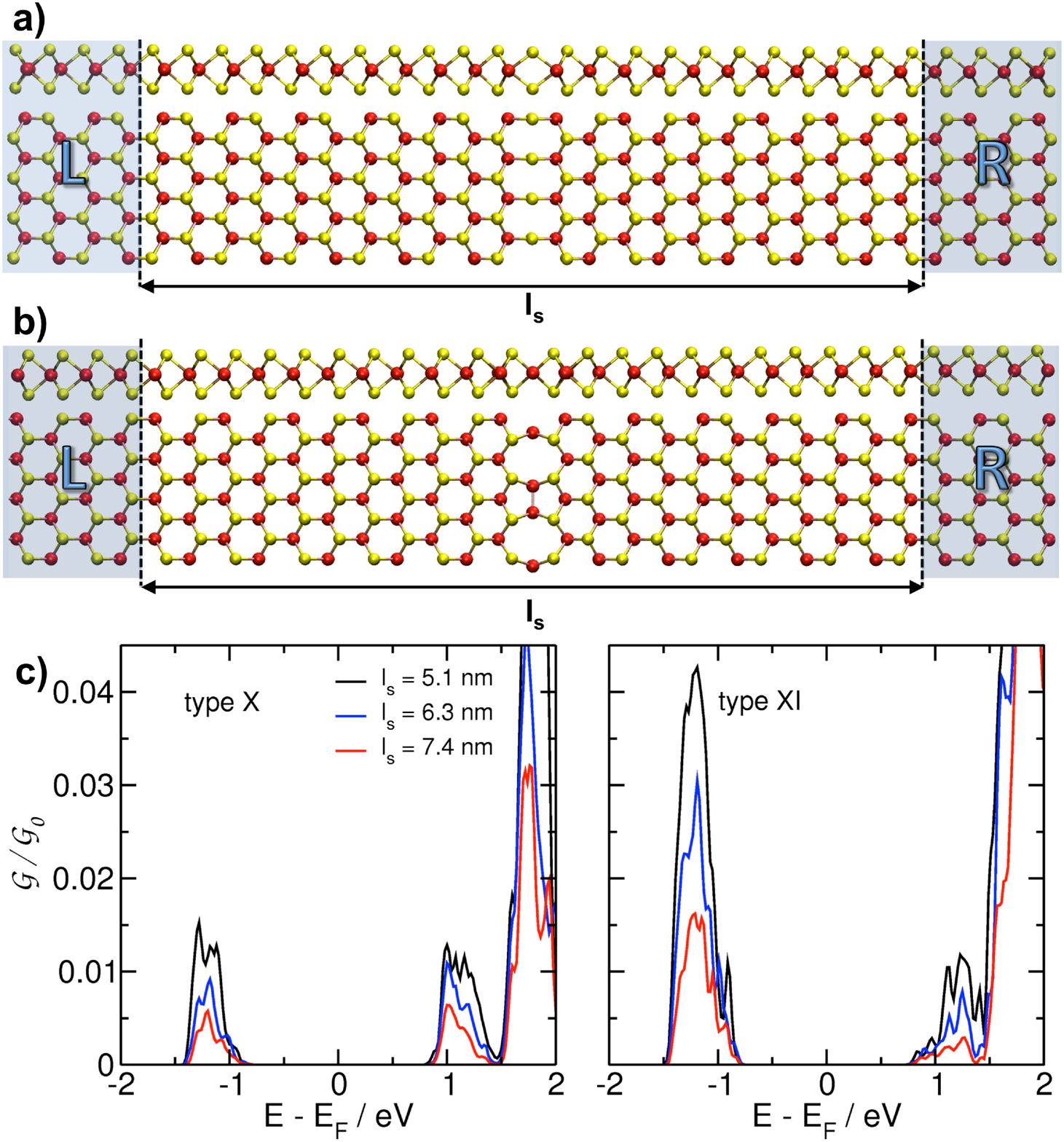}
\caption{\label{fig:6}(Color Online) Schematic representation of the electronic transport in the MoS$_2$-ML with line defects X (a) and XI (b), and the corresponding electron conductance as function of the length of the scattering region, l$_s$ (c). l$_s$ in (a) and (b) correspond to the length of 6.3 nm. Labels as in Figure~\ref{fig:2}.}
\end{center}
\end{figure}

We have also studied the conductance across line defects (X, XI) with respect to the length of the scattering region l$_s$ (see Figure~\ref{fig:6}).
In this case, the structure is periodic along the line defects but in the perpendicular direction there is a mirror symmetry, which should be considered in the transport simulations and the choice of the electrodes.
Here, our electrodes are still perfect MoS$_2$-MLs, but represent mirror images with respect to each other.
Therefore, we have decided to vary the length of l$_s$ and investigate its influence on the transport properties.

Our results show that the conductance at about +/-1.5~eV from the Fermi level reduces with increasing the channel length, however the band gap does not change and no open channels are present close to the E$_F$.

\section{Conclusion}

In summary, we investigated the coherent electron transport through MoS$_2$-ML with various defects on the basis of the Green's functions technique and the DFTB method.
The presence of local defects leads to the occurrence of mid-gap states in semiconducting MoS$_2$-ML.
These states are localized and act as scattering centers.
Our transport calculations show that single-atomic vacancies can significantly reduce the average conductance.
The decrease of conductance depends on the type and concentration of the defects, and, surprisingly, on the transport direction.
We find significant anisotropy of electron transfer in MoS$_2$-ML with grain boundaries.
Since structural and electronic properties of layered semiconducting TMD are comparable, we expect similar effects to occur in other defective TMD-ML.
Our results indicate that structural defects and grain boundaries are principal contributors to the electronic transport properties of MoS$_2$ monolayers, thus rationalizing the large variation of electronic conductivity in different samples.

\section{Acknowledgements}
This work was supported by Deutsche Forschungsgemeinschaft (HE 3543/18-1), the Office of Naval Research Global (Award No N62909-13-1-N222) the European Commission (FP7-PEOPLE-2009-IAPP QUASINANO, GA 251149, and FP7-PEOPLE-2012-ITN MoWSeS, GA 317451) and ERC project (INTIF 226639). 

\providecommand*\mcitethebibliography{\thebibliography}
\csname @ifundefined\endcsname{endmcitethebibliography}
  {\let\endmcitethebibliography\endthebibliography}{}

\end{document}